\documentclass[oldversion]{aa}  
\usepackage{graphicx}
\usepackage{txfonts}
\usepackage{dblfloatfix}
\usepackage{fixltx2e}
\usepackage{natbib}
\usepackage{enumitem}
\usepackage[caption = false]{subfig}
\usepackage{float}
\usepackage{url}
\bibpunct{(}{)}{;}{a}{}{,}

\def\xmm{{\it XMM-Newton~}}
\def\chandra{{\it Chandra~}}

\def\lunits{$\rm erg\,s^{-1}$~}
\def\funits{$\rm erg\,cm^{-2}\,s^{-1}$~}

\begin{document}

\title{Compton-thick active galactic nuclei from the 7 Ms observation in the Chandra Deep Field South}


   \author{A. Corral
     \inst{1,2}
     \and
     I. Georgantopoulos\inst{1}
\and
A. Akylas\inst{1}
\and
P. Ranalli\inst{3,4} 
}

  \titlerunning{CT AGN in the CDF-S}
  \authorrunning{A. Corral et al.}

   \institute{Institute for Astronomy, Astrophysics, Space Applications, and Remote Sensing (IAASARS), National Observatory of Athens, 15236 Penteli, Greece    
\and
Instituto de F\'isica de Cantabria (IFCA), CSIC-UC, Avenida de los Castros s/n, 39005, Santander, Spain
\and
Lund Observatory, Department of Astronomy and Theoretical Physics, Box 43, SE-22100, Lund, Sweden
\and
Combient Mix AB, PO Box 2150, 40313, Gothenburg, Sweden
}

   \date{Received ; accepted }

   \abstract{We present the X-ray spectroscopic study of the
     Compton-thick (CT) active galactic nuclei (AGN) population within
     the \chandra Deep Field South (CDF-S) by using the deepest X-ray
     observation to date, the \chandra 7 Ms observation of the
     CDF-S. We combined an optimized version of our automated
     selection technique and a Bayesian Monte Carlo Markov Chains
     (MCMC) spectral fitting procedure, to develop a method to
     pinpoint and then characterize candidate CT AGN as less model
     dependent and/or data-quality dependent as possible. To obtain reliable
     automated spectral fits, we only considered the sources detected
     in the hard (2-8 keV) band from the CDF-S 2 Ms catalog with
     either spectroscopic or photometric redshifts available for 259
     sources. Instead of using our spectral analysis to decide if an
     AGN is CT, we derived the posterior probability for the column
     density, and then we used it to assign a probability of a source
     being CT. We also tested how the model-dependence of the spectral
     analysis, and the spectral data quality, could affect our results
     by using simulations. We finally derived the number density of CT
     AGN by taking into account the probabilities of our sources being
     CT and the results from the simulations. Our results are in
     agreement with X-ray background synthesis models, which postulate
     a moderate fraction (25\%) of CT objects among the obscured AGN
     population.}

   \maketitle
%

\section{Introduction} 
X-ray surveys are very efficient in detecting active galactic nuclei
(AGN) \citep{xue11,brandt15}. X-rays can penetrate large amounts of
dust and gas by piercing through the obscuring screen that hides the
nucleus. Moreover, as X-rays are largely uncontaminated by the host
galaxy emission, they can more readily probe lower luminosity AGN than
observations at longer wavelengths. The most sensitive observations of
the X-ray Universe, the 7 Ms survey in the \chandra Deep Field South
(CDF-S), reveal a surface density of a few tens of thousand AGN per
square degree \citep{luo17}. For comparison purposes, the quasar (QSO)
sky density obtained through color surveys (e.g., \citealt{croom09})
is only a couple of hundred per square degree.

However, even the extremely efficient X-ray surveys have difficulties
detecting the most highly obscured AGN. Obscured AGN are a key
ingredient in models for galaxy formation and evolution. Among them,
Compton-thick AGN are the most difficult to detect and characterize
because of the huge amount of intervening material obscuring their
intrinsic emission. At the same time, in order to derive a complete
census of Compton-thick AGN, and then to determine the possible
dependence of obscuration on their intrinsic properties and their
evolution, it is of vital importance to constrain these models. X-ray
observations, often complemented with observations in other
wavelengths, are still the best way to carry out the least-biased
studies of this type of sources (see \citealt{hickox18} for a recent
review).

Compton-thick (CT) AGN have column densities (N$_{\rm H}$) higher than
$10^{24}$ $\rm cm^{-2}$, so that Compton scattering becomes an
important contributor to the attenuation of X-rays besides
photoelectric absorption, which is the main absorption mechanism at
lower column densities. These extreme column densities render the CT
AGN about two orders of magnitude fainter in the 2-10 keV band while
the harder energies above 10 keV pass relatively unscathed from the
obscuring screen. Therefore, a very effective way to search for the
most heavily obscured AGN is by using the very hard surveys performed
by {\it SWIFT/BAT} (see \citealt{ricci15,akylas16}) and {\it NUSTAR}
\citep{alexander13,harrison16}. An alternative route is to use very
deep X-ray surveys in the softer 2-10 keV band that manage to detect
highly obscured AGN at very faint fluxes due to their much higher
sensitivity. Previous attempts to detect highly obscured and CT AGN in
this band include the works of \citet{georg13}, \citet{brightman14},
\citet{buch14}, and \citet{liu17}, all of which use \chandra and \xmm
observations in the CDF-S.

In this paper we exploit the deepest X-ray survey ever obtained, the 7
Ms CDF-S. This work differs from previous studies on the X-ray
analysis of the CDF-S in that we do not separate CT AGN from non-CT
AGN, but we compute the probability of a source being CT and use this
probability to derive our results. Additionally, even though we use
7Ms X-ray spectral data, we use the source catalog obtained by
\citet{luo08} from the 2 Ms \chandra observations. This is because we
require our sources to have a sufficient number of counts so that the
characterization of a source as CT is as unambiguous as
possible. Given the minimum number of source counts reported in
\citet{luo08}, we expect our spectral data to have at least $\sim$ 30
counts in the hard (2-8 keV) band.

\section{Chandra Deep Field South}
\begin{figure}[h]
\centering
   \includegraphics[angle=0,width=8cm]{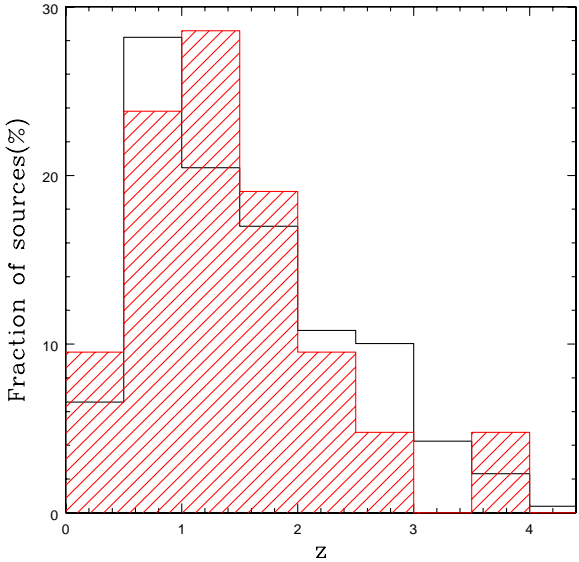} 
 \caption{Redshift distribution for full hard sample (solid histogram), and CT AGN with probabilities $>$ 90\% described in Sect.~\ref{ctagn} (shaded histogram).}
\label{zdistb}
\end{figure}

The CDF-S is the deepest X-ray survey to date, covering an
area of 484.2 arcmin$^{2}$. The most recent catalog of X-ray sources
within the CDF-S was produced by using 102 observations with a total
exposure time of $\sim$ 7 Ms \citep{luo17}. Here we use all these
publicly available observations but, to be able to carry out reliable
spectral fits for all the sources in our sample, we restricted our
analysis to the sources detected in the hard band within the 2 Ms
catalog presented in \citet{luo08}. Therefore, our sensitivity limit
corresponds to 1.3$\times$10$^{-16}$ \funits in the hard band
\citep{luo08}. The hard sample is composed of 265 sources, and either
spectroscopic (181 sources) or photometric (78 sources) redshifts
are available for 259 of them \citep{hsu14}. The redshift distribution is
presented in Fig.\ref{zdistb}.

\begin{figure}[h]
\centering
   \includegraphics[angle=0,width=8cm]{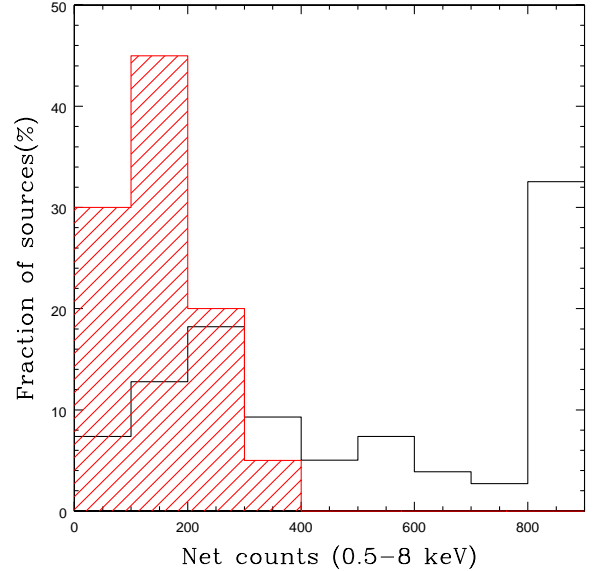} 
 \caption{Net (background subtracted) full (0.5-8 keV) band counts for full hard sample (solid histogram), and CT AGN with probabilities $>$ 90\% described in Sect.~\ref{ctagn} (shaded histogram). The last bin in the solid histogram corresponds to the total percentage of sources with more than 800 counts.}
\label{cdistb}
\end{figure}

We reduced all the \chandra survey data in a uniform manner as
described in \citet{laird09} with the {\tt CIAO} data analysis
software version 4.8. We used the {\tt SPECEXTRACT} script in the {\tt
  CIAO} package to extract source spectra (with an extraction radius
increasing as a function of the off-axis angle to enclose 90\% of the
point spread function, PSF, at 1.5 keV), as well as response and
auxiliary matrices, for each individual observation. The spectral data
from each observation were then merged to create a single source
spectrum, response, and auxiliary matrices for each source using the
{\tt FTOOL} tasks {\tt MATHPHA}, {\tt ADDRMF}, and {\tt ADDARF,}
respectively. Background spectra were extracted in five different
source-free regions for each observation. Then, for each source, the
closest background region among these five was selected and then
merged for each source by taking each source detection (or
non-detection) for each observation into account. Since sources near
the edges of the field may not be present in all individual
observations because of variations in the aim points and roll angles,
the final exposure times range between 400 ks and 6.6 Ms. The net
(background subtracted) counts distribution in the full (0.5-8 keV)
band is presented in Fig.\ref{cdistb}.

\section{Automated spectral analysis}
\label{autofit}
By comparing the different results obtained throughout the years for
the same sources
\citep{norman02,tozzi06,comastri11,georg13,brightman14,buch14,liu17},
it is clear that CT classification is model and data-quality
dependent. Although the sample used in this work is rather small, our
aim is to use it to develop an automated spectral selection method as
less model-dependent as possible that may be applied to large samples
of X-ray spectra spanning a wide range in data quality, such as the
samples from surveys already carried out by \chandra and \xmm, and also
the ones expected to be carried out by the upcoming X-ray missions
{\it SRG/eROSITA}\footnote{http://www.mpe.mpg.de/eROSITA} and {\it
  Athena}\footnote{ttps://www.cosmos.esa.int/web/athena}. The main
characteristic of our method is that we do not classify sources as CT
or not, but we derive the probability of a source being CT
\citep{akylas16}. Therefore in this work, we consider that \emph{an
  AGN is a CT candidate if the resulting probability of the AGN being
  CT is larger than zero.}

We used {\tt Xspec} v12.9 \citep{arnaud} to carry out the spectral
analysis. We selected Cash statistics applied to spectra binned to 1
counts/bin, which allow us to obtain reliable spectral results even
for low count data.

Given the data quality among our sample, we selected the following set
of models, listed in increasing complexity order, to be applied to our data:
\begin{itemize}
\item{An absorbed power-law plus a Gaussian emission line: {\tt
    Xspec:zwabs*zpo+zgaus}. The Gaussian component is intended to
  represent the Fe K$\alpha$ emission line, the most often observed
  emission line in AGN X-ray spectra. In CT AGN, the equivalent width
  (EW) of this line is expected to be very high, sometimes over 1 keV.}
\item{A double power-law plus a Gaussian line: {\tt
    zwabs*zpo+zpo+zgaus}. The unabsorbed power-law can represent
  either soft-scattering of the primary (hard) power-law emission, or
  transmitted emission in the case of a partial covering absorber.}
\item{A modified power-law by a toroidal-shaped absorber plus a
  soft-scattered component: {\tt torus+zpo}, where {\tt torus}
  correspond to the model described in \citet{brightman11}. This model
  is a more appropriate model for our highly absorbed sources since it
  consistently takes into account photoelectric absorption,
  fluorescence line emission (Fe K$\alpha$), and Compton scattering. The
  parameters of this model are, besides the column density, photon
  index, and normalization, the torus opening angle and its inclination
  with respect to the observer. We fixed this angles to 60 and 80
  degrees, respectively (see the following discussion).}
\end{itemize}

Out of the three spectral models described above, the last one is the
more physically-motivated one. However, high data quality is needed to
fit both the opening and inclination angles at the same time. Fixing
the opening angle to 60 degrees and the inclination angle to 80
degrees is a good approximation in the case of modeling highly
absorbed and CT AGN spectra, but it is not always a good choice for
less absorbed sources. For moderately to low absorbed sources, the
derived column densities depend more strongly on the inclination angle
than for highly absorbed ones \citep{lanzuisi18}. Whereas simple
models, like the first two, have been proven to be a good
representation of the spectral shape up to column densities of
10$^{23}$ cm$^{-2}$ \citep{liu17}. It is important to remember that,
up to this point, we are not trying to derive the best-fit model
parameters but to characterize the spectral shape well enough to be
able to pinpoint highly absorbed sources.

Therefore, we proceeded with the following two steps: we applied the
first two models to all of our sample, and then we only applied the
torus model to sources displaying high absorption features in their
spectra. We looked for these features by using the spectral-fitting
results of the first two models (see Sect.~\ref{sims} and
\citealt{corral14}).

\subsection{Optimization of the automated spectral analysis}
\label{sims}

Instead of fitting a highly-absorbed-AGN oriented model (the torus
model) to all of our sources, we made use of the automated selection
technique for highly obscured AGN presented in \citet{corral14}. This
is a fast, less model-dependent, and reliable way to pinpoint heavily
obscured sources without the need to apply complex models that could
be biased toward certain kinds of AGN. This method uses very simple
spectral models (an absorbed power-law: {\tt zwabs*zpo+zgaus}; and a
double power-law: {\tt zwabs*zpo+zpo+zgaus}) to select sources as
highly absorbed candidates by pinpointing signatures of obscuration.

To define a region in the best-fit spectral parameters space so that
all CT sources would be selected, we refined the selection technique
presented in \citet{corral14} by using simulations. We simulated each
source and background in our sample five times by varying the column
density from 10$^{22}$ to 10$^{25}$cm$^{-2}$, while also preserving
their fluxes and redshifts. For column densities below
5$\times$10$^{23}$cm$^{-2}$ we used the {\tt Xspec} model {\tt plcabs}
\citep{yaqoob97} to reproduce the absorbed power law, whereas for CT
column densities, we used the {\tt mytorus} self-consistent model
\citep{murphy09}. In both cases, an additional soft-scattered
power-law component was also included with a maximum scattered
fraction of 10\% with respect to the primary emission. We then applied
our simple models to the simulated data. In order to down-select all
the CT sources, the best-fit parameter for the considered models had
to fulfill at least one of the following selection criteria: the
measured column density is in the CT regime ($>$ 10$^{24}$cm$^{-2}$);
the power-law photon index is flat ($<$ 1.4), either in the hard 2-10
keV band or in the rest-frame hard band, and the column density value
is lager than 5$\times$10$^{23}$cm$^{-2}$; the power-law photon
index is flat ($<$ 1.4), either in the hard 2-10 keV band or in the
rest-frame hard band, and the Fe K$\alpha$ emission line EW is larger
than 500 eV; and for sources with the lowest number of counts in the 2-8
keV band, and thus a limiting reliability of the spectral fits, we
only require that the power-law photon index is flat ($<$ 1.4), either
in the hard 2-10 keV band or in the rest-frame hard band.

By using these criteria, we ensure that all the CT AGN are selected,
albeit including a significant percentage of low ($\leq$ 6\%; N$_{\rm H}$
$<$ 10$^{23}$ cm$^{-2}$) and highly, but not CT, absorbed sources
($\leq$ 17\%; 10$^{23}$ cm$^{-2}$ $<$ N$_{\rm H}$ $<$ 10$^{24}$ cm$^{-2}$).

\subsection{Data quality effects}
\label{dataq}

\begin{figure}[h]
\centering
   \includegraphics[angle=0,width=0.5\textwidth]{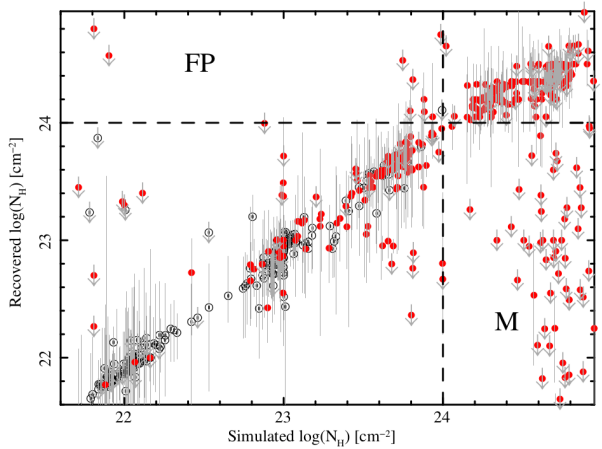} 
   \caption{Simulated column densities ({\tt Xspec: zwabs,plcabs,mytorus}) versus best-fit (recovered) column densities (\citealt{brightman11} torus model). Filled (red) circles correspond to highly absorbed candidates (see Sect.~\ref{sims}). FP: false positives. M: missing CT AGN.}
\label{simplot}
\end{figure}

We also used the simulations described in the previous section to
quantify how the data quality could limit our results. We applied the
torus model described in Sect.~\ref{autofit} to all of the simulated
data, thus obtaining the values displayed in Fig.~\ref{simplot}. There
are two clear problematic regions in this plot labeled FP (false
positives) and M (missing CT AGN).

The most populated among the problematic regions in
Fig.~\ref{simplot} is region M, which corresponds to simulated CT AGN
that are not identified as such when fit by the torus model. All
these simulations are associated with spectra with less than 100
counts in the hard (2-8 keV) band. Given the spectral counts
distribution in our sample, this implies that we could miss up to 14\%
of CT candidates in the case of spectra with less than 100 counts,
that is, we could miss six CT candidates of our actual sample due to low data quality alone.

The second problematic, but much less populated, region in
Fig.~\ref{simplot} is the FP region, which corresponds to sources
misidentified as CT candidates when fit by the torus model. These
simulations also correspond to spectra with less than 100 counts in
the hard band, and they could amount to up to 2\% (only one source in
the actual sample) of misclassified CT candidates.

By combining the results from Sect.~\ref{sims} and the ones presented
in this section, we can estimate the number of missing and/or
misclassified CT candidates in the following analyses (see
Sect.~\ref{lognsim}).

\section{Bayesian CT probabilities}
\label{ctagn}

By applying the selection technique described in Sect.~\ref{sims} to
our actual sample, we ended up with 59 highly absorbed
candidates. According to our simulations, most of these sources should
be at least highly absorbed (N$_{\rm H}$ $>$ 10$^{23}$ cm$^{-2}$), and
all of the CT AGN in our sample should be within these 59 candidates.

As mentioned at the end of Sect.~\ref{autofit}, once we identified the
sources most likely to be CT AGN within our sample, we applied a more
appropriate model to them in order to obtain our final spectral
results as follows: the torus model described in \citet{brightman11}
and a second power law that accounts for soft scattered emission. From
the spectral results of this model, we confirm that most ($\sim$ 90\%)
of our candidates are in fact highly absorbed AGN.

\begin{figure*}[!htb]
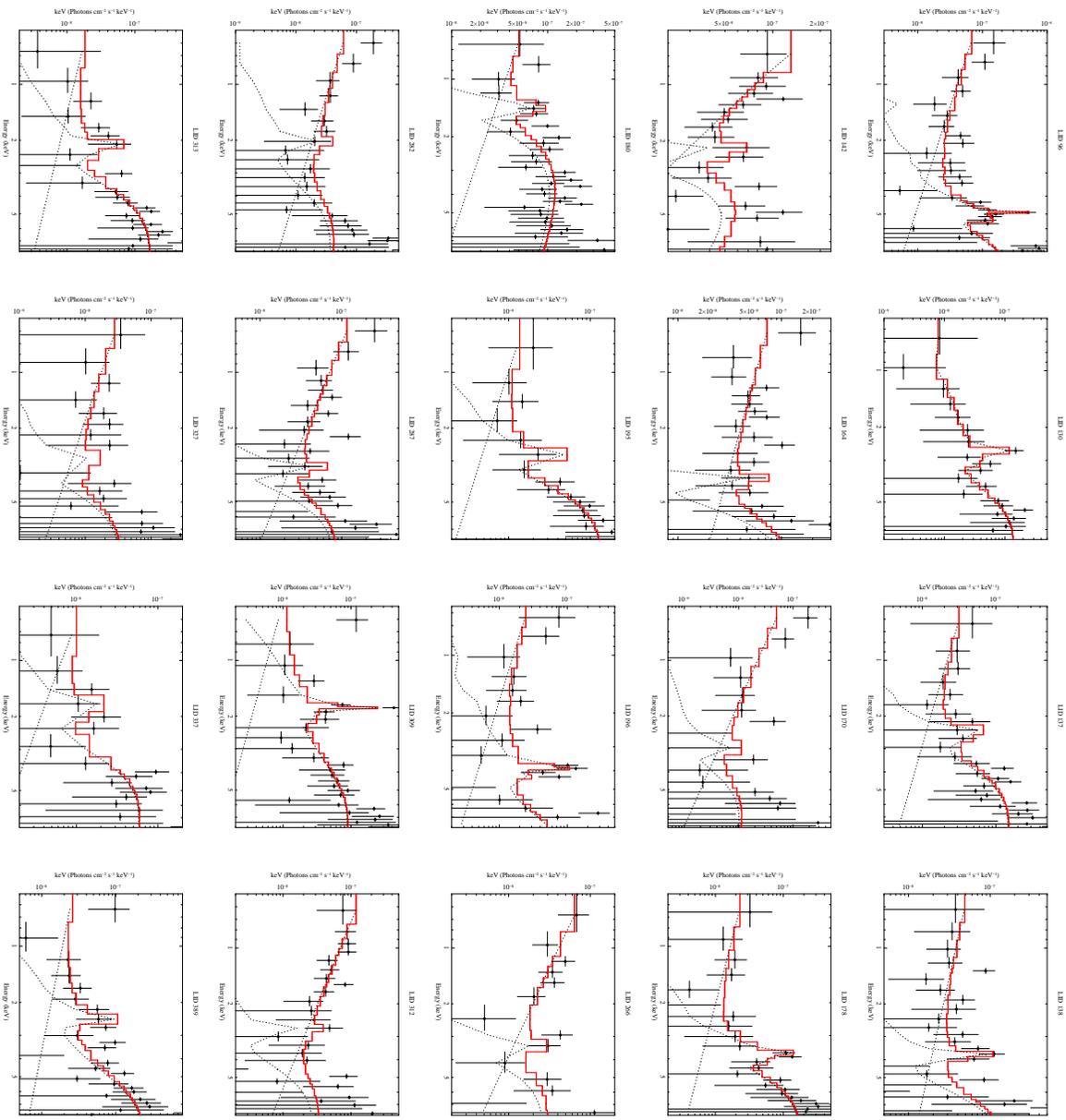

  \begin{tabular}{cccc}
\includegraphics[angle=-90,width=0.2\textwidth]{96_euf.ps} & \includegraphics[angle=-90,width=0.2\textwidth]{130_euf.ps} & \includegraphics[angle=-90,width=0.2\textwidth]{137_euf.ps} & \includegraphics[angle=-90,width=0.2\textwidth]{138_euf.ps}\\
\includegraphics[angle=-90,width=0.2\textwidth]{142_euf.ps} & \includegraphics[angle=-90,width=0.2\textwidth]{164_euf.ps} & \includegraphics[angle=-90,width=0.2\textwidth]{170_euf.ps} & \includegraphics[angle=-90,width=0.2\textwidth]{178_euf.ps}\\
\includegraphics[angle=-90,width=0.2\textwidth]{180_euf.ps} & \includegraphics[angle=-90,width=0.2\textwidth]{195_euf.ps} & \includegraphics[angle=-90,width=0.2\textwidth]{196_euf.ps} & \includegraphics[angle=-90,width=0.2\textwidth]{266_euf.ps}\\
\includegraphics[angle=-90,width=0.2\textwidth]{282_euf.ps} & \includegraphics[angle=-90,width=0.2\textwidth]{287_euf.ps} & \includegraphics[angle=-90,width=0.2\textwidth]{309_euf.ps} & \includegraphics[angle=-90,width=0.2\textwidth]{312_euf.ps}\\
\includegraphics[angle=-90,width=0.2\textwidth]{313_euf.ps} & \includegraphics[angle=-90,width=0.2\textwidth]{327_euf.ps} & \includegraphics[angle=-90,width=0.2\textwidth]{337_euf.ps} & \includegraphics[angle=-90,width=0.2\textwidth]{389_euf.ps}
  \end{tabular}
\caption{X-ray unfolded spectra for  candidate CT AGN with probabilities $>$ 90\%.}
  \label{newct}
\end{figure*}

Instead of basing a CT classification on the best-fit column density,
we adopted the Bayesian approach described in \citet{akylas16}. This
approach is based in the derivation of the probability distributions
for each variable spectral parameter by using Monte Carlo Markov
Chains (MCMC) and the Goodman-Weare algorithm. In this way, instead of
classifying a source as CT or not, we assign a probability of the CT
source by integrating the probability distribution of the column
density above 10$^{24}$ cm$^{-2}$. In applying this method, we find
that 36, out of the initial 59 sources, have a probability higher than
zero of being CT (P$_{\rm CT} >$ 0); 20 of them have probabilities $>$
90\% (see Fig.\ref{newct} and Table~\ref{chandrafits}). Taking the
probabilities for all of the 36 candidates into account, the effective
number of CT AGN in our sample is 27.

It is important to remember that our intention is not to classify
sources as CT AGN nor to derive the actual column densities of our
sources, but to derive the probability of being CT (P$_{\rm
  CT}$). Therefore, the comparison presented in the last column of
Table~\ref{chandrafits} must only be considered qualitatively and not
as a direct comparison (see Sect.~\ref{compare}). We tested that by
applying different models such as {\tt plcabs, mytorus}, and/or by
adding reflection ({\tt Xspec: pexmon}), whereas changing the
resulting values for the best-fit column densities did not
significantly affect the resulting integrated probabilities over
10$^{24}$ cm$^{-2}$.  
\begin{table*}[h]
\caption{Spectral fits results for candidate CT AGN} 
\label{chandrafits}
\centering
\begin{small}
$$
\begin{tabular}{cccccccccc}
\hline\hline
LID  & z  & $\rm log\,N_{\rm H}$  &  $\Gamma$ &   P1/P2 &   Flux  & $\rm log\,L_{\rm X}$       & $\rm log\,L_{\rm Xunabs}$ & CT probability & PC\\
 &    &      cm$^{-2}$   &    &       & $10^{-15}$ \funits & \lunits & \lunits & (\%) &\\ 
(1) & (2) & (3) & (4) & (5) & (6) & (7) & (8) & (9) & (10)\\
 \hline
96 & 0.310 & $24.14^{+0.64}_{-0.13}$ & $2.02^{+0.33}_{-0.36}$ & 0.004 & 1.48 & 41.4 & 42.9 & 96.7 & CT \\ \\
130 & 1.327 & $24.47^{+0.44}_{-0.30}$ & $2.00^{+0.35}_{-0.49}$ & $<0.001$ & 1.38 & 42.3 & 44.2 & 98.4 & nCT  \\ \\
137 & 1.544 & $24.13^{+0.34}_{-0.12}$ & $1.57^{+0.52}_{-0.11}$ & 0.027 & 1.40 & 42.47 & 43.7 & 95.0 & CT \\ \\
138 & 0.738 & $24.43^{+1.50}_{-0.30}$ & $1.68^{+0.61}_{-0.21}$ & 0.014 & 0.92 & 41.9 & 43.4 & 100.0 & noCT \\ \\
142 & 1.86  & $24.05^{+1.24}_{-0.05}$ & $1.92^{+0.46}_{-0.40}$ & 0.15 & 0.68 & 42.8 & 43.6 & 94.5 & noCT \\ \\
164 & 0.729 & $24.20^{+1.63}_{-0.07}$ & $1.45^{+0.63}_{-0.10}$ & 0.09 & 0.87 & 41.9 & 42.9 & 100.0 & noCT \\ \\
170 & 1.22  & $24.41^{+1.34}_{-0.15}$ & $2.17^{+0.36}_{-0.22}$ & 0.04 & 0.12 & 41.7 & 43.0 & 99.4 & nCT \\ \\
178 & 0.738 & $24.07^{+0.36}_{-0.10}$ & $1.72^{+0.46}_{-0.18}$ & 0.007 & 1.32 & 41.9 & 43.2 & 92.9 & CT \\ \\
180 & 3.660 & $24.02^{+0.14}_{-0.10}$ & $1.68^{+0.28}_{-0.27}$ & 0.006 & 1.28 & 43.3 & 44.2 & 90.0 & CT\\ \\
195 & 1.41 & $24.23^{+0.54}_{-0.14}$ & $1.71^{+0.46}_{-0.24}$ & 0.008 & 1.00 & 42.1 & 43.7 & 98.3 & CT\\ \\
196 & 0.670 & $24.35^{+0.43}_{-0.14}$ & $1.98^{+0.48}_{-0.40}$ & 0.004 & 0.70 & 41.58 & 43.34 & 99.9 & CT\\ \\
266 & 0.954 & $24.07^{+1.82}_{-0.11}$ & $2.00^{+0.51}_{-0.25}$ & 0.093 & 0.33 & 41.8 & 42.8 & 99.4 & noCT  \\ \\
282 & 2.223 & $24.47^{+0.55}_{-0.09}$ & $2.19^{+0.32}_{-0.52}$ & 0.040 & 0.36 & 42.7 & 43.9 & 99.9 & CT \\ \\
287 & 1.037 & $24.21^{+1.26}_{-0.07}$ & $1.94^{+0.43}_{-0.38}$ & 0.052 & 0.77 & 42.2 & 43.4 & 97.2 & CT \\ \\
309 & 2.578 & $25.10^{+0.88}_{-0.27}$ & $1.75^{+0.41}_{-0.14}$ & 0.001 & 0.98 & 42.8 & 44.6 & 100.0 & CT\\ \\
312 & 1.309 & $24.26^{+1.65}_{-0.32}$ & $2.10^{+0.28}_{-0.37}$ & 0.085 & 0.36 & 42.3 & 43.2 & 100.0 & noCT\\ \\
313 & 1.608 & $24.25^{+0.20}_{-0.14}$ & $1.62^{+0.48}_{-0.31}$ & 0.012 & 1.51 & 42.4 & 43.9 & 99.3 & CT\\ \\
327 & 1.096 & $24.12^{+1.70}_{-0.36}$ & $1.92^{+0.49}_{-0.47}$ & 0.082 & 0.18 & 41.7 & 42.7 & 99.6 & \ldots \\ \\
337 & 2.470 & $24.48^{+0.80}_{-0.16}$ & $1.95^{+0.0.25}_{-0.49}$ & 0.002 & 0.55 & 42.3 & 44.2 & 99.6 & CT\\ \\
389 & 1.74 & $24.59^{+1.26}_{-0.14}$ & $1.67^{+0.47}_{-0.23}$ & 0.006 & 1.68 & 42.6 & 44.3 & 100.0 & nCT\\ \\
\hline \hline
\end{tabular}
$$
\begin{list}{}{}
\item{The columns are: (1) \chandra ID from the \citet{luo10}
  catalog. (2) Redshift (two decimal and three decimal digits denote
  photometric and spectroscopic redshifts, respectively). (3)
  Intrinsic column density. (4) Photon index ($^{f}$ corresponds to
  fixed photon index). (5) Ratio between the scattered and the torus
  components. (6) Observed flux in the 2-10 keV band. (7) Observed
  luminosity in the 2-10 keV band. (8) Unabsorbed luminosity in the
  2-10 keV band.(9) Probability of the source being
  Compton-thick. (10) Previous classification from \citet{buch14}, \citet{brightman14}, and/or \citet{liu17}:
  Compton-thick (CT), near-CT ({\rm $N_{\rm H} > 5\times10^{23}$}, nCT), not a CT (noCT).}
\end{list}
\end{small}
\end{table*}

\section{Discussion} 
\subsection{Comparison with previous results}
\label{compare}
\citet{brightman14,buch14,liu17} performed systematic studies of CT
AGN within the CDF-S by using \chandra data, finding nine, eight, and
ten candidate CT AGN, respectively. However, their CT classifications
are not consistent with each other in many cases, being that 15 AGN is
the combined number of candidates from those
works. \citet{brightman14,buch14} used 4Ms CDF-S spectral data but not
exactly the same sources, whereas \citet{liu17} used 7Ms data but only
presented the spectral analysis for the brighter sources in that
catalog.

In this work, we derived CT probabilities $>$ 90\% for only 10 out of
the 15 previously reported CT AGN included in
\citet{brightman14,buch14,liu17} (see Table~\ref{chandrafits}). For
the remaining 5 AGN previously classified as CT, we find them only
moderately to highly absorbed, except for source 180 which we find to
be near-CT (probability $\sim$ 89\%). More importantly, we find nine
additional CT AGN with probabilities > 90\% that were not classified
as such in any of those works. Three of them were previously
classified as near-CT. Another one, source 327 (\citealt{luo08} ID),
was not included in any of the samples used in
\citet{brightman14,buch14,liu17}. For the remaining five, either the
column density, the power-law photon index, or both, were fixed during
the previous spectral fits, which could explain the different
results. In the case of sources 138, 266, and 312, the differences
could be also due to the improved spectral quality in the 7 Ms data,
since these sources are not included in the \citet{liu17} sample. New
CT candidates are mainly due to lower number of counts in
\citet{brightman14,buch14}, which used 4Ms data, or because the
sources are not included in either of the
\citet{brightman14,buch14,liu17} samples.

\subsection{Comparison with X-ray background synthesis models}
\label{lognsim}
We compared our CT number count distribution with the most recent
cosmic X-ray background (CXB) synthesis models of \citet{akylas12} and
\citet{ueda14}. In our case, we weighted each of our candidate CT by
the probability of them being CT (see \citealt{akylas16}), so that the
number integral number count, N($>$S), of sources per unit sky area
with flux higher than S is defined as follows:

\begin{centering}
\begin{equation}
N(>S_{j}) = \sum_{i=1}^{i=j}\frac{P_{\rm CT}}{\Omega_{i}}
,\end{equation}
\end{centering}

where we summed all the sources with fluxes
S$_{i}>$S$_{j}$ at each bin, and $\Omega_{i}$ represents the sky coverage as a
function of flux from the 2Ms survey\citep{luo08}.

\begin{figure}[h]
\centering
   \includegraphics[angle=0,width=9cm]{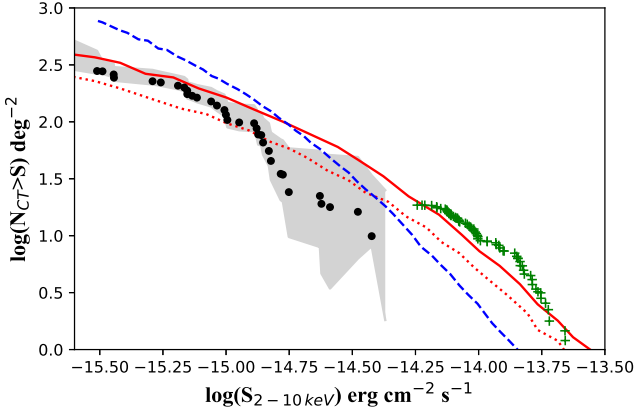} 
   \caption{Number count distribution of CT AGN (circles) in CDF-S (this work) and for COSMOS-Legacy data in
     \citet{lanzuisi18} (crosses). Solid and dotted lines correspond
     to the model predictions presented in \citet{akylas12} for a CT
     fraction of 15\% plus 5\% reflection, and for a CT fraction of 25\%,
     respectively. The dashed line corresponds to the model in
     \citet{ueda14}.}
\label{lognlogs}
\end{figure}

Instead of computing the errors in each bin, we estimated a confidence
interval by performing simulations. According to the results from
Sect.~\ref{dataq}, we could be missing up to 6 CT candidates, and one
of our candidates could have been wrongly selected. We also know that
these seven sources have to fulfill the selection criteria described
in Sect.~\ref{sims}, that is, they must be among the 59 highly
absorbed candidates with derived P$_{\rm CT}$ = 0. Besides, their
spectra must have less than 100 counts in the hard band. Taking all of
this into account, we simulated 10000 realizations in which we added
up to six sources (simulated following the fluxes of the highly
absorbed candidates with less than 100 counts, and with random P$_{\rm
  CT}$), and randomly removed one of the actual CT candidates. We
considered our confidence interval to be the region that encompasses
99.7\% of our simulated number count distributions (gray area in
Fig.\ref{lognlogs}).

Our results are plotted in Fig.~\ref{lognlogs}. As a comparison, we
plotted the models of \citet{akylas12} for a 15\% CT fraction and 5\%
of the reflected emission, for a 25\% CT fraction, and the model of
\citet{ueda14}, which assumes a large ($\sim$ 50\%) percentage of CT
AGN and only a moderate amount of reflection. One of the main
differences between both models is that the \citet{ueda14} model has
an extra free parameter allowing for differential evolution of the CT
AGN population relative to the population of unobscured and mildly
obscured AGN. We also plotted the results from \citet{lanzuisi18},
which also use \chandra data but from the 3Ms COSMOS-Legacy survey
\citep{civano16}. In that work, they also derived the probability
distribution for the column densities and use them in a similar way as
we did in this work. Our results look more consistent with the model
proposed in \citet{akylas12} for a 25\% percentage of CT AGN, a
smaller value than the one presented in \citet{lanzuisi18}. They
report a number density of CT AGN which increases from 30-55\% with
redshift. Because of the relative small size of our sample, we cannot
test this evolution with redshift in our case. A very recent work that
also makes use of the CDF-S 7Ms data, although a different spectral
analysis was carried out in that case, reports a moderate fraction of
CT when computing their number counts \citep{li19}, which is
consistent with our results.  Although smaller fractions and/or CXB
models with higher amounts of reflection are also consistent within
errors, local studies point to small amounts of reflection
\citep{georg19}.

\section{Conclusions}
We used the deepest X-ray observation to date, the 7 Ms \chandra
observation of the CDF-S, to search for CT AGN in the hard (2-8 keV)
selected sample presented in \citet{luo08}, which is based on a
shallower 2Ms observation. In this way, we were able to improve
previous spectral analyzes and to better constrain the intrinsic
column densities of the sources in our sample. Moreover, by making use
of simulations, we estimate that among X-ray spectra with less than
100 counts in the hard band (2-8 keV), X-ray analyses could be missing
$\sim$ 14\% of the CT AGN just because of the data quality.

To optimize the automated spectral analysis and to quantify
data-quality effects, we applied an automated selection technique to
select highly absorbed candidates, and then we applied a Bayesian
method to compute the probability of a source being CT. We find 36 CT
candidates with probabilities larger than zero, and 20 with
probabilities $>$ 90\%. Nine of them are classified as CT for the
first time thanks mostly to the deeper 7Ms data. Our results are in
good agreement with previous results, although we do not confirm the
previous CT classification of five sources. We find that one source is
near-CT, and that the other four look to be only moderately absorbed.
  
We used the computed probabilities to derive the number count
distribution of CT AGN in the CDF-S. By comparing our findings with
the CXB synthesis models in \citet{ueda14,akylas12}, our results favor
the one presented in \citet{akylas12} assuming a percentage of 25\% CT
AGN.
\begin{acknowledgements}   
We thank the referee for his/her helpful comments and suggestions.The
{\it Chandra} data were taken from the Chandra Data Archive at the
Chandra X-ray Center. AC acknowledges funding from ESA under the
PRODEX project and financial support through grant
AYA2015-64346-C2-1-P (MINECO/FEDER). AC is supported by a ``Juan de la
Cierva Incorporaci\'on'' postdoctoral contract from Ministerio de
Ciencia, Innovaci\'on y Universidades (Spain).
\end{acknowledgements}

\bibliographystyle{aa}
\bibliography{corral_cdfs_7ms}

\end{document}